# C-sanitized: a privacy model for document redaction and sanitization

David Sánchez, Montserrat Batet

Departament d'Enginyeria Informàtica i Matemàtiques,
UNESCO Chair in Data Privacy, Universitat Rovira i Virgili,
Av. Països Catalans 26, E-43007 Tarragona, Spain

Corresponding author: David Sánchez; E-mail: david.sanchez@urv.cat

**Abstract**

Within the current context of Information Societies, large amounts of information are daily exchanged and/or released. The sensitive nature of much of this information causes a serious privacy threat when documents are uncontrollably made available to untrusted third parties. In such cases, appropriate data protection measures should be undertaken by the responsible organization, especially under the umbrella of current legislations on data privacy. To do so, human experts are usually requested to redact or sanitize document contents. To relieve this burdensome task, this paper presents a privacy model for document redaction/sanitization, which offers several advantages over other models available in the literature. Based on the well-established foundations of data semantics and the information theory, our model provides a framework to develop and implement automated and inherently semantic redaction/sanitization tools. Moreover, contrary to ad-hoc redaction methods, our proposal provides *a priori* privacy guarantees which can be intuitively defined according to current legislations on data privacy. Empirical tests performed within the context of several use cases illustrate the applicability of our model and its ability to mimic the reasoning of human sanitizers.

# C-sanitized: a privacy model for document redaction and sanitization

## Introduction

Information Societies require governments, public administrations and enterprises to exchange and release large amounts of sensitive information for various reasons. For example, the Freedom of Information Act (U.S. Department of Justice, 2013) forces the government to release confidential information. In some countries, healthcare organizations require to provide medical records to insurance companies in response to Worker's Compensation and Motor Vehicle Accident claims (Gordon, 2013). Because of the confidential nature of much of that information, responsible organizations should take appropriate protection measures to guarantee the fundamental right to privacy of individuals, thus fulfilling current legislations (Department of Health and Human Services, 2000; Health Privacy Project, 2013; The European Parliament and the Council of the EU, 1995).

To guarantee individuals' privacy, that is, the ability to seclude the sensitive information of an individual from the others, data protection/sanitization obfuscates or removes sensitive information. Traditionally, data protection has been performed manually, in a process by which a group of experts rely on their knowledge to detect and obfuscate the entities that, under their opinion, may disclose sensitive information by means of *semantic inferences* (e.g., treatments or drugs that may reveal sensitive diseases, readings that may suggest political preferences, habits that can be related to religion or sexual orientations, etc.) (Gordon, 2013). Semantics are in fact crucial in data privacy because they define the way by which humans (sanitizers, data analysts and also potential attackers) understand and manage information. However, because of the enormous amount of information that is daily disclosed (e.g., the US Department of Energy's OpenNet initiative (U.S. Department of Energy, 2012) requires to sanitize millions of documents yearly), there is a need for automatic methods and tools in order to avoid the bottleneck and cost of manual sanitization (Dorr, Phillips, Phansalkar, Sims, & Hurdle, 2006; Gordon, 2013).

The computer science community, under the umbrella of the *privacy-preserving data publishing* research field (Fung, Wang, Chen, & Yu, 2010), has proposed a plethora of protection methods. Most of them rely on formal *privacy models* to attain a *predefined* notion of privacy (Drechsler, 2011). Privacy models have been very popular in recent years because, in comparison to ad-hoc approaches in which the level of protection is empirically evaluated *a posteriori* for a specific dataset (Hundepool et al., 2013), models offer *a priori* privacy guarantees. Moreover, widely accepted privacy models provide a *de facto* standard to develop specific privacy-preserving methods, which can be objectively compared and evaluated under homogenous conditions. Finally, the *a priori* guarantees offered by these models fit better with current legislations and privacy guidelines (Department for a Healthy New York, 2013; Department of Health and Human Services, 2000; Health Privacy Project, 2013; Legal Information Institute, 2013; Nicholson & Smith, 2007; Terry & Francis, 2007; The European Parliament and the Council of the EU, 1995), which also give *a priori* rules on what and how much sensitive information should be protected. For example, the HIPAA "Safe Harbor" rules (Department of Health and Human Services, 2000) state that "all geographic subdivisions smaller than a state" should be removed from patient Electronic Healthcare Records, whereas US state and federal privacy legislations (Health Privacy Project, 2013) specify that any information about HIV status or substance abuse should be removed prior releasing records for secondary use. More broadly, EU regulations forbid to process personal data that may reveal racial or ethnic origin, political opinions, religious or philosophical beliefs and any data concerning health or sex life (The European Parliament and the Council of the EU, 1995).

According to the type of data to release (e.g., relational databases, transactional data or unstructured documents), the research community has proposed different privacy models. As it will be discussed in the next section, researchers have put a lot of effort into developing privacy models for structured data. Unfortunately, privacy models aiming at protecting unstructured documents, which are the focus of this paper and the most usual way of information exchange between actors, are much scarcer.

**Related works**

Privacy protection is usually referred to in the literature as the minimization of the probability of *identity* and/or *attribute disclosure*. The former covers the protection of information that can re-identify an individual, and is usually referred to as *anonymity*, whereas the latter covers the protection of *confidential* data (while identities may be left in clear). Privacy models usually aim at balancing the trade-off between data privacy and data utility (Domingo-Ferrer, 2008), which would range between a perfect protection in which data is removed or encrypted while its utility is completely destroyed, and a direct publication of unprotected data, thus providing maximum utility but with no privacy protection at all.

The first theoretical privacy model was proposed in (Dalenius, 1977), which stated that the access of a potential attacker to the protected data of an individual should not increase the attacker's knowledge of the confidential information of that individual. This is a very strong assumption and it has been demonstrated that it is not practically feasible (Drechsler, 2011) when background knowledge is available, which is the most usual scenario.

Recent works have development more practical models. Within the area of structured databases (also known as microdata), in which released data consist in records detailing a set of attributes of individuals (e.g., census data), *k*-anonymity (Sweeney, 2002) is the most well-known privacy model. It is based on the classification of record attributes as:
  i) *Identifiers*: these univocally identify an individual and are systematically removed prior releasing (e.g., ID cards, SS numbers).
  ii) *Quasi-identifiers*: attributes whose combination of values may unequivocally disclose identities (e.g., age+job+place of birth). They have to be sanitized prior releasing.
  iii) *Confidential attributes* (e.g., salaries): these are left in clear form to retain the analytical utility of the published data.

As the model name suggests, *k*-anonymity focuses on ensuring *anonymity*, that is, the impossibility of re-identifying any individual from the published data, while leaving anonymous confidential data in clear. The privacy guarantee specifies that, for a data set to satisfy *k*-anonymity, any record in the set must be indistinguishable with regard to the values of the quasi-identifier attributes from, at least, k-1 other records in the same set (identifying attributes are assumed to be removed). Practical implementations of the *k*-anonymity model usually group records in sets of cardinality *k* and replace them by a prototypical tuple (Domingo-Ferrer, 2008; Martínez, Sánchez, & Valls, 2013). Although *k*-anonymity helps to minimize identity disclosure, it may not protect against confidential attribute disclosure. This is the case of a group of *k*-anonymous records that share the same confidential value (e.g., patients being or not AIDS-positive): even though an attacker would not be able to identify a particular individual from the group of $k$ records, he will learn the individual's confidential attribute because all the records in the group share the same value. Several authors have proposed extensions of the $k$-anonymity model to *partially* fix that issue: *l*-diversity (Machanavajjhala, Kifer, Gehrke, & Venkitasubramaniam, 2007), which requires a minimum variance between confidential attribute values of each group of $k$ records, and *t*-closeness (Li & Li, 2007), which requires that grouped confidential attribute values follow a similar distribution to that of the whole dataset.

From a different point of view, the more recent $\varepsilon$-differential privacy model (Dwork, 2006) seeks to limit the knowledge gain provided by the protected data with regard to any individual in the input data set. Thus, it focuses on avoiding the disclosure of *confidential* information, rather than on anonymity. It requires the protected data to be insensitive to modifications of *one* input record with a probability depending on $\varepsilon$. In this manner, an attacker would not be able to unequivocally disclose the information of a specific individual with a probability depending on $\varepsilon$. To achieve this guarantee, most practical works add noise to the data in a magnitude so that the protected output becomes insensitive (according to the $\varepsilon$ parameter) to a modification of an input record. Since the amount of noise required to fulfill this guarantee for the whole dataset is usually very high (Soria-Comas, Domingo-Ferrer, Sánchez, & Martínez, 2014), thus compromising the utility of the output, most works apply $\varepsilon$-differential privacy in an interactive setting. In this scenario, a trusted party holds the database and answers queries sent by users with sanitized responses. This introduces limitations on both the number and type of supported query responses.

The previous models rely on the structure of data (i.e., records with univalued attributes) to define, achieve and evaluate privacy guarantees. Much less attention has been paid to the definition of models for less structured data sets (e.g., lists of transactions) and completely unstructured data (e.g., plain text documents).

For transactional data (like query logs (Batet, Erola, Sánchez, & Castellà-Roca, 2013)), where individuals' information is represented by lists of items with variable and usually large cardinality, we find the $k^m$-anonymity model (Terrovitis, Mamoulis, & Kalnis, 2008). It considers that any combination with any cardinality of items in the transaction list may disclose the individual; thus, any item(s) in the list may act as quasi-identifier(s). The $k^m$-anonymity model ensures that the dataset is $k$-anonymous in front of an adversary with background knowledge of up to $m$ items of any transaction. Similarly to the basic $k$-anonymity model, $k^m$-anonymity may fail when background knowledge allows inferring causalities between items (Anandan & Clifton, 2011).

Plain text documents (e.g., e-mails, sales or company descriptions, medical visit outcomes, etc.) are even more challenging. Their lack of structure prevents from defining *a priori* sets of identifying/quasi-identifying attributes. Like for transactional data, any combination of textual entities may produce disclosure. Thus, the barrier between anonymity and confidentially is diluted because i) confidential pieces of text may also enable re-identification, and ii) the privacy needs related to the document exchange may differ from one scenario to another: *anonymity* (e.g., to preserve the identity of the company described in a document) or *confidentiality* (e.g., to hide the fact that an individual suffers from a sensitive condition). In order to preserve the privacy in textual documents, two tasks are usually performed: i) detection of textual terms that may cause disclosure of sensitive information, which is usually tackled manually and ii) removal or obfuscation of those entities. The community refers to the act of removing or blacking-out sensitive entities as *redaction*, whereas *sanitization* consists on their obfuscation by replacing them with appropriate generalizations (e.g., AIDS->disease) (Bier, Chow, P. Golle, T. H. King, & Staddon, 2009). Obviously, the latter approach is more desirable, since it better preserves the utility of the protected output. Moreover, in document redaction, the existence of blacked-out parts in the released document can raise the awareness of the document's sensitivity in front of potential attackers (Bier et al., 2009), whereas sanitization gives no clues on entities' sensitivity. Unfortunately, document sanitization has received less attention than the more harmful redaction (Anandan et al., 2012).

Regarding the protection of barely semantic identifying data in textual documents (e.g., ID numbers, addresses, ages, dates, etc.), there exist automatic methods that exploit their regular structure in order to detect them by means of rules, patterns or trained classifiers (Meystre, Friedlin, South, Shen, & Samore, 2010). As stated in current legislations like the HIPAA (Department of Health and Human Services, 2000), there are scenarios in which these identifiers should be directly removed/redacted in order to preserve the anonymity, as done within the context of structured databases. After that, the remaining textual entities, which lack a regular structure and may act as quasi-identifiers or refer to confidential information (e.g., sensitive diseases or topics like religion, race or sexuality), should be considered in order to minimize the disclosure risk inherent to semantic inferences. This can be done by enforcing a privacy model.

Privacy models focusing on plain text documents are scarce and non-widely adopted. Moreover, they have a limited scope since they only focus on the protection of sensitive terms, which are assumed to be detected/defined *a priori*. First, we find two models that adapt the notion of *k*-anonymity in structured databases: *K*-safety (Chakaravarthy, Gupta, Roy, & Mohania, 2008) and *K*-confusability (Cumby & Ghani, 2011). Both approaches assume the availability of a large enough and, ideally, homogenous collection of documents (e.g., contents of e-mails of employers from a company). They require each sensitive entity contained in each document of the collection to be indistinguishable from, at least, K-1 other entities present in the collection. For the *K*-safety model, this should be also fulfilled with regard to the contexts with which sensitive entities usually co-occur (e.g., *names* and *birth places* for *persons*). In both cases, the authors generalize terms or contexts, so that they become less diverse and, hence, indistinguishable enough. As drawbacks, documents cannot be sanitized individually, and, due to the need to generalize heterogeneous terms to a common abstraction, a high loss of utility will happen if document contents are not homogenous.

Finally, in (Anandan et al., 2012) the authors present a more practical model named *t*-plausibility, which allows to sanitize textual documents individually. The model relies on generalizing sensitive entities and states that a document holds *t*-plausibility if, at least, *t* documents can be derived from the protected document by specializing sanitized entities. In other words, the protected document generalizes, at least, *t* plausible documents obtained by combining specializations of sanitized terms. To do so, it is required a detailed knowledge base covering every possible sensitive entity, both to retrieve generalizations used to sanitize sensitive terms and to count the number of plausible versions that can be obtained by means of specializations. This limits the accuracy of the method according to the suitability of the knowledge base with respect to the contents of the document to be protected. The authors also note that, with the *t*-plausibility model, some entities could be generalized more than others, which would result in some entities being worse protected than others. To tackle this problem, the authors also propose a more practical model (uniform *t*-plausibility) that tends to ensure that all entities are generalized uniformly. In this case, they seek for an ad-hoc balance between an optimal *t*-plausibility and a sanitization in which they generalize all sensitive entities to the same level of abstraction. As drawback for both models, it is noted that setting the *t*-plausibility level is not intuitive for end users and that one can hardly predict the results, because they depend on the document size, the number of sensitive entities and the amount of generalizations/specializations modeled in the knowledge base. Finally, these approaches evaluate and sanitize textual terms independently, neglecting the fact that, due to semantic inferences (Anandan & Clifton, 2011; D. Sánchez, Batet, & Viejo, 2013c), combinations of several terms (e.g., symptoms and medical procedures) may unequivocally disclose sensitive information (e.g., a disease).

**Contributions and plan**

In this paper we propose a new privacy model that tackles the limitations of available privacy models focusing on unstructured textual documents. Its goal is to mimic and, hence, automatize the reasoning of human sanitizers with regard to semantic inferences, disclosure analysis and protection of textual documents. To achieve that, our proposal relies on an assessment and quantification of the data semantics that human experts usually consider in document sanitization (Bier et al., 2009; Gordon, 2013). Our proposal provides the following contributions over the state of the art:

- In comparison with available models (Anandan et al., 2012; Cumby & Ghani, 2011), which assume that *all* risky terms (sensitive entities or related terms) have been identified a priori, our proposal automates both the *detection* of terms that can disclose sensitive data via semantic inferences and their *protection*. This relieves human sanitizers from manually identifying related terms, which has been identified as one of the most difficult and time-consuming challenges (Bier et al., 2009; Gordon, 2013). To do so, our model considers, as human sanitizers do, the semantic relationships by which terms or combinations of terms appearing in a document would disclose sensitive information via semantic inferences. Moreover, our proposal automates both the redaction and sanitization of risky terms. Hence, it facilitates the design of automatic methods that cover the whole sanitization/redaction task.
- Our approach offers a flexible and intuitive way of setting the privacy requirements in specific scenarios, by stating *what* is considered sensitive and *up to which degree* it should be hidden. Contrary to models like *t*-plausibility (Anandan et al., 2012), it offers a clear and *a priori* privacy guarantee on how sensitive entities will be protected, which does not depend on the document size or the number of sensitive entities. Moreover, the privacy criterion can be set using intuitive linguistic labels rather than abstract numbers. As a result, the model can be easily and straightforwardly instantiated according to current legislations and regulations on data privacy (Department for a Healthy New York, 2013; Department of Health and Human Services, 2000; Health Privacy Project, 2013; Legal Information Institute, 2013; Terry & Francis, 2007; The European Parliament and the Council of the EU, 1995), whose rules are also expressed linguistically (e.g., disease names, locations, etc.). Finally, our model supports both anonymity and confidentiality, thus enabling its application in scenarios in which identities should be hidden (Department of Health and Human Services, 2000) and in which confidential attributes should be protected (Health Privacy Project, 2013).

- Contrary to privacy models derived from *k*-anonymity (Cumby & Ghani, 2011; Chakaravarthy et al., 2008), our proposal supports the sanitization of individual documents, which is very convenient when managing heterogeneous sets of documents or in environments in which input documents are produced sequentially (e.g., e-mails from an individual's mail account).
- Contrary to *t*-plausibility (Anandan et al., 2012), our model does not require a knowledge base to protect the input document, even though its availability will contribute to improve the utility of the sanitized output.

The enforcement of our model builds on the foundations of the Information Theory to ensure that the offered privacy guarantees are coherent with how textual information is used and how their semantics are interpreted by humans (either sanitizers or potential attackers). Moreover, it is sustained by previous research (D. Sánchez, Batet, & Viejo, 2013a; D. Sánchez, Batet, et al., 2013c; D. Sánchez, Batet, & Viejo, 2014; David Sánchez, Batet, & Viejo, 2012), in which the suitability of its theoretical basis was empirically studied and evaluated.

The rest of the paper is organized as follows. The second section presents and formalizes our privacy model, detailing its privacy guarantees and the theoretical principles that enable its enforcement. The third section discusses some relevant aspects related to the model implementation and proposes a simple and scalable algorithm for document sanitization. The fourth section evaluates and discusses the empirical results obtained by the model implementation for several case studies. The final section depicts the conclusions and presents some lines of future research.

## A privacy model for document redaction/sanitization

In general, information disclosure happens when an attacker is able to unequivocally discover some sensitive information (e.g., identities, confidential data) by analyzing the released data and exploiting his own knowledge. In order to ensure a robust privacy protection, this should be done in the worst-case scenario: attackers who know or have access to all the domain knowledge (Hundepool et al., 2013).

More specifically, within the context of unstructured data protection, our goal is to mimic human judgments with regard to document sanitization, so that human experts can be relieved from such time-consuming task. As stated in the introduction, because human sanitizers analyze data from a semantic perspective and evaluate disclosure according to semantic inferences (Bier et al., 2009; Gordon, 2013), we interpret disclosure in terms of semantics of data. Informally, given the whole domain knowledge, this can be assessed by answering to this question: does a released term or a combination of terms in a document allow to unequivocally infer (via semantic inference) and, thus, disclose a sensitive entity?; by unequivocal we refer to the fact that there is no semantic ambiguity in the correct inference given the domain knowledge.

Our privacy model, named *c-sanitized*, aims to avoid such *unequivocal* semantic inference/disclosure of sensitive entities in unstructured documents, while retaining *some* of their semantics in the protected document. In this manner, the utility of the protected document is partially preserved, which is mandatory for the purpose of information exchange/release. The basis of our model can be defined as follows.

**Definition 1**. *c-sanitized*: given an input document *D*, the whole domain knowledge *K* and a sensitive entity *c* to be protected, we say that *D'* is the *c-sanitized* version of *D* if *D'* does not contain any term *t* or group of terms *T* that individually or in aggregate can unequivocally disclose *c* by exploiting *K*.

The sensitive entity *c* should correspond to the most general entity of the topic that must be protected prior releasing the document (e.g., a kind of disease such as *cancer*). Our model relies on the evaluation of the disclosure risk that terms present in *D* cause with regard to *c* given the domain knowledge *K*. In cases of disclosure, an enforcement of our model should take appropriate sanitization measures (e.g., removal/generalization) to those terms *t* or group of terms *T* that individually or in aggregate can unequivocally disclose *c*. For example if *c* refers to a sensitive disease (e.g., *AIDS*), *t* may refer to unequivocal agents (e.g., *HIV*) and *T* to groups of elements that, even though individually may seem innocuous, might be univocally associated to the disease when interpreted in aggregate (e.g., *immune system + sexual transmission + influenza-like disease*). From a practical perspective, our model instantiation can be performed in accordance with privacy legislations stating sensitive entities or topics in order to avoid disclosing, for example, HIV related data within a health record (Health Privacy Project, 2013); in this example, the document would be *HIV-sanitized*.

From the perspective of the formal privacy notions commonly used in the literature, if *c* refers to confidential information (e.g., a sensitive condition such as *HIV*), the protection process will be implicitly focused on the protection of *confidentiality/attribute disclosure*. On the contrary, if *c* refers to identifying information (e.g., census data or geographical divisions such as *Los Angeles*), the protection process will be implicitly focused on protecting *anonymity/identity disclosure*. This provides a natural adaptation to different privacy needs and scenarios. If several topics should be sanitized (e.g., medical conditions, but also census-related data like state names), several instantiations of the model with different entities *c* can be iteratively applied over the same document (e.g., *HIV-sanitized*, *Los Angeles-sanitized*) so that each sanitization step is made in coherence with the topic defined by each *c*. In such cases, we generalize our model for lists of sensitive entities (e.g., {*HIV-Los Angeles*}-*sanitized*), as follows.

**Definition 2**. *C-sanitized*: given an input document *D*, the whole domain knowledge *K* and a set of sensitive entities *C* to be protected, we say that *D'* is the *C-sanitized* version of *D* if *D'* does not contain any term *t* or group of terms *T* that that individually or in aggregate can unequivocally disclose *any* entity in *C* by exploiting *K*.

The next section details how the model can be enforced by quantifying the entity disclosure caused by semantic inferences from an information theoretic perspective.

**An information theoretic enforcement**

The cornerstone of the proposed model is how to assess whether *c* or those entities in *C* are or are not unequivocally disclosed by the presence of a term or group of terms in *D*. As discussed above, this kind of inferences is enabled by the semantics inherent to the released data that the domain knowledge *K* enable (Chow, Golle, & Staddon, 2008). That is, if according to *K*, terms in *D* are closely semantically related to entities in *C*, the risk of disclosure will be high.

However, since semantics is an inherently human and qualitative feature, the detection of possible inferences and the quantification of the resulting disclosure are not trivial. To tackle this problem, we adopt an information theoretic quantification of data semantics and, thus, of disclosure risk. The basic idea is that the semantics encompassed by a term appearing in a context can be quantified by the *amount of information* it provides, that is, its *Information Content* (IC).

The notion of IC has been extensively used in the past to quantify term semantics and to compare the resemblance between terms in the area of *semantic similarity* (Resnik, 1995; D. Sánchez, Batet, & Isern, 2011). More specifically, the IC of a textual term *t* and, thus, the semantics encompassed by *t*, can be computed as the inverse of its probability of occurrence in corpora (which represents the knowledge *K*).

$$IC(t) = -\log p(t) \tag{1}$$

Ideally, if corpora are large and heterogeneous enough to reflect the information distribution at a social scale, resulting IC values will be a faithful representation of term's semantics (Cilibrasi & Vitányi, 2006) as they are understood and used by humans in their communicative acts. In fact, these human actors can be either expert sanitizers, who exploit their knowledge to minimize disclosure risks, authorized readers of the sanitized document, for whom the document's utility directly depends on the semantics encompassed by the terms appearing in it (Martínez et al., 2013), or potential attackers, who can exploit the revealed semantics and available knowledge to disclose sensitive data. Thus, a privacy model that considers the whole semantics of terms as they are understood by society will implement a realistic notion of the practical privacy risks and of data utility. In fact, empirical experiments conducted in (D. Sánchez, Batet, et al., 2013a, 2013c; D. Sánchez et al., 2014) illustrated the direct relationship between the amount of semantics encompassed by textual terms (i.e., their informativeness) and their disclosure risk and sensitivity.

Notice that by applying the notion of IC to the sensitive entity *c*, it turns that $IC(c)$ is measuring the semantics of *c* and, thus, the amount of *sensitive* information that should be protected. Since the revelation of this *amount of information* in the output document is what unequivocally provides the semantics of *c* and, thus, discloses *c*, our model and the notion of unequivocal disclosure (Definition 2) can be enforced in terms of Information Theory as follows.

**Definition 3**. *C-sanitized*: given an input document *D,* the whole domain knowledge *K* and a set of sensitive entities *C* to be protected, we say that *D'* is the *C-sanitized* version of *D* if, for all *c* in *C, D'* does not contain any term *t* or group of terms *T* that, according to *K*, reveal $IC(c)$ information about *c*.

As stated earlier, the disclosure risk or the degree of semantic revelation that terms appearing in a document produce with respect to a sensitive entity depends on the closeness of the semantic relationships between them (e.g., a sensitive disease and its unequivocal symptoms). In terms of information, the amount of *c*'s semantics, that is, $IC(c)$, provided/revealed by terms *t* or groups of terms *T* appearing in the document, can be measured according to *Mutual Information* (MI) between two variables (D. Sánchez, Batet, et al., 2013c). The instantiation of MI for two specific outcomes, that is, the term/group of terms *t*/*T* and the sensitive entity *c*, corresponds to their *Point-wise Mutual Information* (PMI). PMI measures the amount of information overlap (see some examples in Figures 1-4) between two entities. PMI has been used in the past to measure the semantic similarity between pairs of terms (Bollegala,

Matsuo, & Ishizuka, 2009; D. Sánchez, Batet, Valls, & Gibert, 2010), that is, the amount of semantics that the two terms reciprocally provide of each other. PMI can be computed as the difference between the normalized probability co-occurrence of the two entities, given their join and marginal distributions in corpora (Church & Hanks, 1990):

$$PMI(c;t) = \log \frac{p(c,t)}{p(c)p(t)} \qquad (2)$$

Numerically, PMI is maximal if, according to $K$, $t$ always co-occurs with $c$, resulting in $PMI(c;t)=IC(c)$. This states that $c$ is completely disclosed by $t$ because the semantics of the former can be inferred from the latter, given their usage in society. Thus, our privacy model can be completely enforced in terms of information quantification, as follows.

**Definition 4**. *C-sanitized*: given an input document $D$, the whole domain knowledge $K$ and a set of sensitive entities $C$ to be protected, we say that $D'$ is the *C-sanitized* version of $D$ if, for all $c$ in $C$, $D'$ does not contain any term $t$ or group of terms $T$ so that, according to $K$, $PMI(c;t)=IC(c)$ or $PMI(c;T)=IC(c)$, respectively.

To fulfill Definition 4, those terms $t$ in $D$ whose PMI with regard to $c$ is equal to the IC of $c$, should be sanitized/redacted from the output document $D'$.

As discussed earlier, even though an individual term $t$ may not disclose the semantics of $c$, a combination of several individually innocuous terms $T$ (e.g., combinations of treatments and symptoms) may produce a larger disclosure when considered in aggregate, because the degree of semantic revelation in such cases is the union of the individual revelations caused by each element of the group. To evaluate groups of terms $T$, where $T=\{t_1,...,t_n\}$ without ordering, PMI can be formulated as follows:

$$PMI(c;T) = \log \frac{p(c,t_1,...,t_n)}{p(c)p(t_1,...,t_n)} \qquad (3)$$

In this manner, combinations of terms that fully reveal the semantics of $c$ (that is, that always co-occur with $c$) can be also detected, even though each term of the group may not individually disclose $c$.

**Characterizing the semantics of disclosure**

To better understand the semantics that term (co-)occurrences and IC/PMI scores quantify, and how these enable disclosure, in this section we characterize different types of information disclosure according to the type of semantic relationship between the sensitive entity $c$ and a term $t$ occurring in the document to protect:
- $t$ and $c$ are taxonomically related. Several cases can be distinguished:
  - The term $t$ is a generalization of $c$ (e.g., $c=AIDS$, $t=immune system disease$ or $c=Hepatitis C$, $t=Hepatitis$): since the semantics of $t$ are completely encompassed by its specialization $c$, which inherits the semantic features of its taxonomic ancestors, it can be concluded that $t$ provides a strict *subset* of the information of $c$. This is coherent because, if both the specialization and the generalization appear in the same context, the latter provides no additional information about the former. The graphical representation of the corresponding disclosure in terms of informativeness is shown in Figure 1. Thus, according to the privacy model, it is not necessary to sanitize/redact $t$. In terms of probability, term generalization occurrences must be counted whenever any of the specializations occur (Resnik, 1995). This results in monotonically increasing probabilities as terms become more general, that is, $p(t)>p(c)$ and, thus, $IC(t)<IC(c)$. Moreover, thanks to the coherent counting, we have that $p(c,t)=p(c)$ and, thus, we obtain $PMI(c;t)=\log(p(c,t)/p(t)p(c))=\log((p(c)/p(t)p(c))= \log(1/p(t))=-\log(p(t))=IC(t)$, which corresponds to the greyed area of Figure 1.

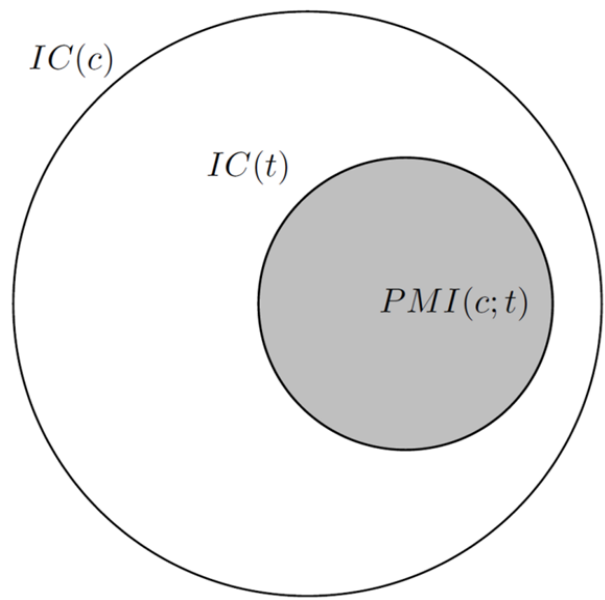

*Figure 1.* Information disclosed (greyed area) of *c* when *t* is a generalization of *c*.

- o  The term *t* is a specialization or a synonym/lexicalization of *c* (e.g., c=*Hepatitis*, t=*Hepatitis C* or c=*immune system disease*, t=*AIDS*, or c=*HIV*, t=*Human Immunodeficiency Virus* or c=*AIDS*, t=*Acquired Immunodeficiency Syndrome*): in case of specialization (the first two examples), that is, the semantics of *c* are encompassed by *t*, *t*'s informativeness is a strict superset of the information provided by *c*. In case of being synonyms (the last two examples), that is, both sharing the same semantics, the informativeness of *c* and *t* are also the same. The disclosure corresponding to these cases is illustrated in Figure 2. Given the previous probability calculus, we have that $PMI(c;t)=IC(c)$ because $p(c,t)=p(t)$ and $p(c)≥p(t)$. Since *t* completely discloses *c* (and even adds more information in case of being a specialization), according to Definition 4, *t* must be redacted/sanitized. Notice that, if a document has been *c-sanitized*, it will be also *c'-sanitized* for all *c'* that are a taxonomic specialization of *c*. For example, the sanitized output will be identical if the document has been *Hepatitis-sanitized* or *{Hepatitis, Hepatitis C}-sanitized*, given that *Hepatitis C* is a specialization of *Hepatitis*. Thus, to avoid redundant calculus, *c* should correspond to the most general entity of each topic that should be protected.

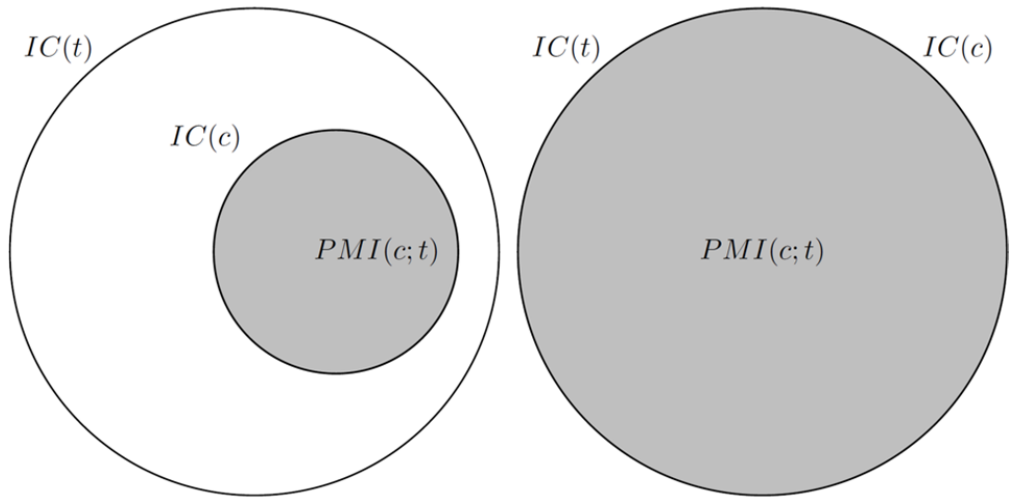

*Figure 2.* On the left: information disclosed of *c* (greyed area) when *t* is a taxonomical specialization of *c*. On the right: information disclosed from *c* (greyed area) when *t* is a synonym of *c*.

- *t* and *c* are non-taxonomically related (e.g., *c=AIDS, t=contaminated blood transfusion* or *c=Hepatitis C, t=liver failure*). A non-taxonomic semantic relationship implies that *c* and *t* share a portion of their semantics, even though they are not strictly encompassed by each other like in the case of taxonomic relationships. Hence, as shown in Figure 3, *t* provides a subset of the information of *c*, whose size is a function of the closeness of the semantic relationship. Thus, *PMI(c;t)≤IC(c)* because *p(c,t)≤p(c)*.

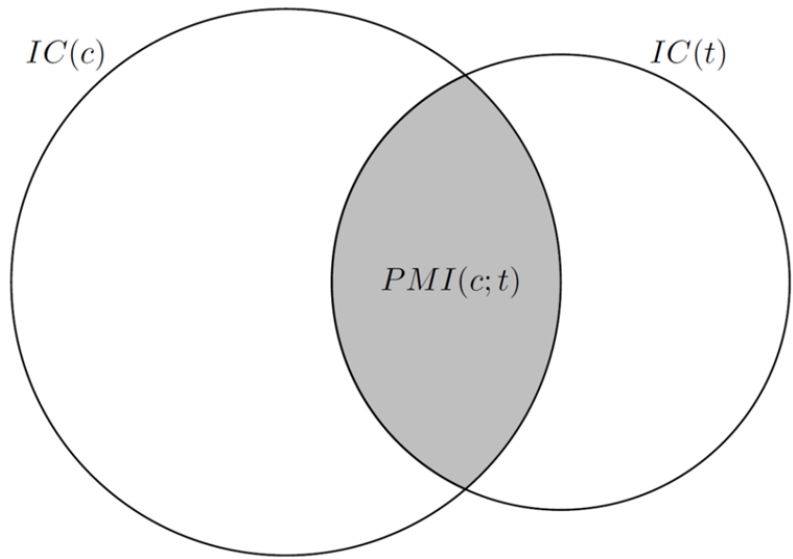

*Figure 3.* Information disclosed of *c* (greyed area) when *t* is non-taxonomically related to *c*.

Moreover, the co-occurrence of several non-identical terms within the same context (constituting the group of terms *T*) may increase the disclosure of the semantics of *c*. As discussed in the previous subsection, this is because each $t_i$ in *T* provides a potentially overlapping but non-identical subset of the semantics of *c*, according to the kind of taxonomic or non-taxonomic relationship that links them (e.g., *c=AIDS*, $t_1$=*immune system disease*, $t_2$=*contaminated blood transfusion*, or *c=Hepatitis C*, $t_1$=*Hepatitis*, $t_2$= *liver failure*). Hence, the evaluation of disclosure should consider the *union* of the information of *c* provided by *all* terms in *T*, as it is shown in Figure 4. This corresponds to *PMI(c;T)*. According to the size of this union, we obtain *PMI(c;T)≤IC(c)* because $p(c,t_1,…,t_n) \leq p(c)$.

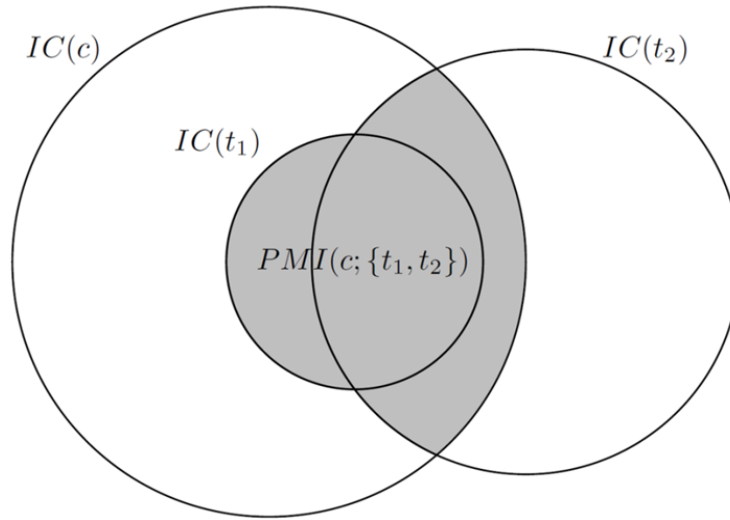

*Figure 4*. Information disclosed of *c* (greyed area) by the co-occurrence of $t_1$ (which is a generalization of *c*) and $t_2$ (which is non-taxonomically related to *c*).

**Preventing disclosure**

According to Definition 4, to avoid disclosing *c*, those *t* or *T* whose PMI with regard to *c* equals *IC(c)* must be masked prior releasing the document. It is quite common in practice to simply remove/black-out risky terms (i.e., *document redaction*) (Chow et al., 2008; D. Sánchez, Batet, et al., 2013c), so that the disclosed information/semantics are annulled. However, this act severely hampers the utility of the output, which closely depends on the preservation of data semantics (Martínez et al., 2013). Moreover, it also raises awareness of the sensitive nature of the document, because of the presence of removed/blacked-out parts (Bier et al., 2009).

Since our privacy model enables retaining a certain degree of semantics/utility without completely disclosing entities in *C*, risky terms *t* (e.g., *AIDS*) are alternatively replaced by an appropriate generalization *g(t)* (e.g., *disease*) in a process referred to as *document sanitization* (Bier et al., 2009). As it is shown in Figure 5, because *g(t)* encompasses the semantics of *t* and thus provides a strict subset of the information of *t*, it also lowers the amount of disclosure of *c*. This allows fulfilling the privacy criterion of Definition 4 while retaining a certain amount of semantics/utility, which can be quantified as *IC(g(t))*.

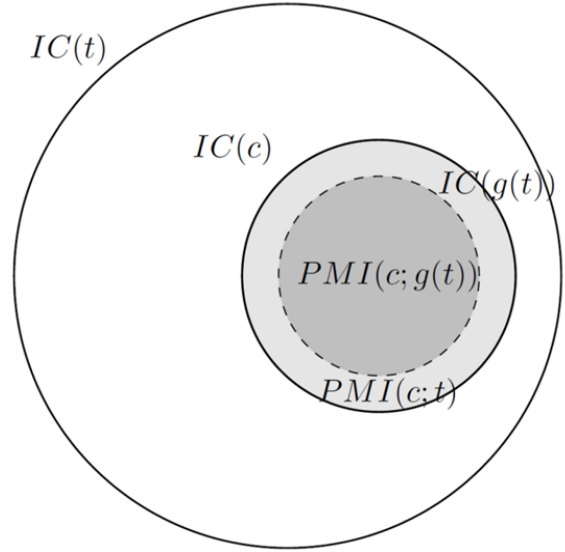

*Figure 5*. Information retained/disclosed (darker greyed area) when sanitizing (i.e., generalizing) the term *t*.

That is, by replacing terms that may cause disclosure of a sensitive entity by appropriate generalizations, we are rendering them equivocal enough (from the perspective of semantic inferences) so that potential attackers will not be able unequivocally discover the entity, which is our goal.

Given a risky term *t*, the optimal generalization *g(t)* from the perspective of utility preservation should retain the maximum amount of semantics (i.e., *IC(g(t))*) while fulfilling the privacy model (i.e., *PMI(c;g(t))<IC(c)*). For individual terms *t*, the selection of the optimal generalization is straightforward. However, when considering groups of terms *T* of any cardinality from *D*, the optimal sanitization (i.e., the combination of generalizations of each $t_i$ in *T* that retains the maximum amount of semantics while fulfilling the privacy model for any group of terms in *D*) is NP-hard. This is also inherent to most privacy models (Anandan et al., 2012; Cumby & Ghani, 2011; Chakaravarthy et al., 2008) because the order in which terms in *D* are evaluated influences the order in which the groups of terms *T* are evaluated and sanitized. Hence, an optimal sanitization would require evaluating *all* possible combinations of terms in *D* of any cardinality and any possible generalization of each $t_i$ in *T*, and picking up the combination of elements that optimizes the result. To render this process practical, different heuristics can be considered according to the priorities of the sanitization process (i.e., data utility, efficiency or disclosure risk). Some discussions on this are presented the next section.

**A parametric definition**

With the proposed model, a *C-sanitized* document will offer the guarantee of non-unequivocal disclosure (i.e., no unambiguous inference) of any entity in *C* given the knowledge *K*. However, this guarantee could be too rigid and may still cause privacy risks because it is quite common to also consider the disclosure of a *significant* amount (but not necessarily all) of the semantics of the sensitive entities as a privacy breach (D. Sánchez, Batet, & Viejo, 2013b).

In order to provide a more flexible solution with regard to different (stricter) privacy needs, we introduce a parameter that lowers the amount of disclosed information bellow the maximum that enables the unequivocal disclosure. This proportionally increases the ambiguity in the attackers' inferences because, according to the knowledge $K$, the less specific disclosed information will enable *several plausible* inferences. Thus, the increased ambiguity produces a lower likelihood that an attacker's belief regarding the sensitive entity is true, because there exist other plausible solutions and there is no way to distinguish them. To do so, we extend Definition 2 as follows:

**Definition 5**. ($\alpha$,C)-*sanitized*: given an input document $D$, the whole domain knowledge $K$ and a set of sensitive entities $C$ to be protected, we say that $D'$ is the ($\alpha$,C)-*sanitized* version of $D$ if $D'$ does not contain any term $t$ or group of terms $T$ that reveal $1/\alpha$ or more of the semantics of *any* entity in $C$ by exploiting $K$.

In terms of Information Theory, Definition 4 can be extended likewise:

**Definition 6**. ($\alpha$,C)-*sanitized*: given an input document $D$, the whole domain knowledge $K$ and a set of sensitive entities $C$ to be protected, we say that $D'$ is the ($\alpha$,C)-*sanitized* version of $D$ if, for all $c$ in $C$, $D'$ does not contain any term $t$ or group of terms $T$ so that, according to $K$, $PMI(c;t) \geq IC(c)/\alpha$ or $PMI(c;T) \geq IC(c)/\alpha$, respectively.

The $\alpha$ parameter is a continuous numerical value in the range $[1..\infty[$. Setting $\alpha=1$, we obtain the initial privacy model (Definitions 2 and 4). Setting $\alpha>1$, the redaction/sanitation process will become stricter, since a smaller amount of disclosure will be allowed. For example, for $\alpha=2$, less than half of the information content (roughly semantics) of any sensitive entity $c$ is allowed to be disclosed. In practice, this adds a proportional degree of uncertainty to the attackers' inferences that lowers the risk of unequivocal disclosure. In any case, how much the information to be released should be lowered to maintain the disclosure risk low enough is a matter of the specific scenario and privacy needs. This issue will be considered in the evaluation section.

## Model implementation

This section discusses the main aspects that should be considered when developing redacting/sanitization methods and tools based on the proposed model, and gives some insights on how to tackle them. Finally, it proposes a simple algorithm that provides a scalable implementation of the model.

### Probability calculus

The information theoretic enforcement of our model extensively relies on the assessment of the informativeness of terms that, in turn, is a function of their probabilities of occurrence/co-occurrence. Thus, the exactness of the probability calculus is crucial to ensure that the model is evaluating a realistic representation of term semantics according to the domain knowledge $K$.

To compute term(s) probabilities for IC/PMI calculus, different resources can be used, which will represent the domain knowledge $K$ considered in the analysis. In the past, many authors relied on *tagged corpora*, which produced accurate probabilities (Resnik, 1995) because terms have been manually associated to their conceptual abstractions and, thus, occurrences are properly disambiguated. Particularly, by following the probability calculus proposed in (Resnik, 1995), in which occurrences of term specializations are also counted as occurrences of their generalizations, it is ensured that probabilities monotonically increase as terms are generalized. However, since corpora are usually tagged manually, the availability of large and representative enough tagged corpora, so that probabilities reflect the information distribution at a social scale (i.e., a realistic representation of social knowledge), is very unlikely. In consequence, even though some tagged corpora (such as the Brown corpus (Francis & Kucera, 1979)) are available, their limited size and scope result in data sparseness, especially for domain-specific terms (e.g., chemical compounds), named-entities (e.g., organization names) or newly minted terms (e.g., a new electronic device) (D. Sánchez et al., 2010). These terms are precisely the usual target of document sanitization due to their high informativeness (D. Sánchez, Batet, et al., 2013a).

To tackle these limitations, it is possible to use large *raw corpora* instead. Contrary to tagged corpora, plain text resources are very common and easily accessible. The paradigmatic example is the Web, in which billions of electronic up-to-date textual documents are available. The use of the Web as corpora for probability calculus provides several advantages: i) its size, heterogeneity and *social* publication paradigm configure a realistic representation of the information distribution at a social scale (Cilibrasi & Vitányi, 2006), ii) Web Search Engines can be used as proxies for probability calculus by querying terms and by evaluating the resulting page count (D. Sánchez, Castellà-Roca, & Viejo, 2013; Turney, 2001), iii) from the perspective of document sanitization, some authors have considered the Web as a reasonable proxy of social knowledge, thus providing a good approximation of the whole knowledge $K$ that attackers may exploit to disclose sensitive data (Staddon, Golle, & Zimmy, 2007). However, probabilities computed from plain text (i.e., by counting the number of explicit occurrences of words) are negatively affected by language ambiguity. This may underestimate the informativeness of a concept (i.e., in case of polysemic words appearing in different contexts with different senses) or overestimate it (i.e., because of synonymy or ellipsis, a number of appearances may not be considered).

One can see that none of the previous approaches is free of flaws. Thus, probability calculus will be usually imperfect, either because of the data sparseness caused by the limited scope of the exploited data source or because of the language ambiguity inherent to plain text corpora. On the one hand, it is a matter of the concrete use case and sanitization needs to prefer one approach to another. On the other hand, the use of the more flexible ($α,C$)-*sanitized* model with $α>1$, which will result in a stricter sanitization, may help to mitigate the imperfections in the disclosure risk assessment caused by i) the underestimation of the probability of $c$, which will result in an overestimated informativeness of the sensitive entity and, thus, in a less strict sanitization, and ii) the overestimation of the probability of co-occurrence of $c$ and $t$, which will result in an underestimation of the amount disclosed information.

**Scalability**

As argued earlier, the optimal sanitization with regards to data utility is NP-hard for the proposed model. This is common to privacy models that try to balance the trade-off between privacy and data utility (Anandan et al., 2012; Cumby & Ghani, 2011). To ensure the practical feasibility and scalability of the *C-sanitized* model in a real scenario, some aspects can be considered during its implementation.

First, divide the input document in several contexts. In general, the closeness of the semantic relationship between terms co-occurring in a document tends to decrease as they are more distant in the text (Lemaire & Denhière, 2006). Hence, it is more likely to discover risky groups of terms within, for example, the same sentence than when terms are located in different paragraphs. By defining textual contexts and analyzing them independently, the number of term combinations to evaluate will significantly decrease. The context length will depend on the tightness of the discourse and also on the privacy needs. Usual textual contexts implemented by existing redacting/sanitization methods cover paragraphs, sentences (D. Sánchez, Batet, et al., 2013b, 2013c) or adjacent words (Anandan & Clifton, 2011), even though the whole document can also be considered in case of very tight discourses (e.g., a patient medical record) (D. Sánchez, Batet, et al., 2013c).

Second, the evaluation of all groups of terms of any length is computationally costly. To mitigate this, it will be preferable to start analyzing individual terms and, then, iteratively add more elements to the group. In this manner, individual terms or smaller groups found to be risky and, thus, sanitized/redacted, would not be evaluated again (D. Sánchez, Batet, et al., 2013c). This will also limit the number of terms that can be combined together and, hence, the number of distinct groups to evaluate.

Finally, an alternative to the evaluation of groups of terms with unbounded cardinality is to define a stricter privacy criterion by using the ($\alpha$,*C*)-*sanitized* model and setting $\alpha>1$. The bigger the $\alpha$ value is, the more likely will individual terms be found to be risky, thus reducing the need to evaluate groups of terms. Even though the omission of groups of terms (or limiting their cardinality) will likely produce less accurate results, it will also enable a highly scalable sanitization, in which the algorithm can solve the problem in linear time with regard to the number of terms in the document.

**Exploiting knowledge bases**

In order to implement a utility-preserving sanitization of sensitive terms (instead of simply removing them), a knowledge base (KB) organized in a taxonomic way is needed to retrieve the term generalizations (e.g., AIDS->immunodeficiency->disease) that will be used as replacements. A suitable KB for document sanitization should ideally provide a high recall, so that all or, at least, most of the potentially sensitive terms found in the input document are covered in the KB. Those sensitive terms that cannot be found in the KB would be removed, with the subsequent utility loss. Moreover, a fine grained taxonomic structure will be also interesting, so that the information loss derived from each generalization step can be minimized. By having fine grained generalizations available for each term, its sanitization can better approximate the optimal situation, that is, the term generalization that, while maximizing the amount of given information, fulfills the privacy guarantee of our model.

A general-purpose ontology such as WordNet, which covers thousands of general concepts and has been translated to many languages, is well suited as a generic solution. In fact, it has already been used by several sanitization methods (Anandan et al., 2012; Cumby & Ghani, 2011). However, WordNet offers a limited coverage of named entities or recently minted terms. In this case, other more dynamic knowledge sources like ODP (the largest human-edited directory service, with more than 1 million categories) or YAGO (a knowledge base created from the structure of categories and articles of Wikipedia) may be more appropriate.

For concrete sanitization scenarios (e.g., a healthcare organization needs to protect a patient record before providing it to an insurance company), domain specific KBs may be preferable. For example, in the medical domain, large and detailed biomedical KBs like SNOMED-CT, which models diseases, substances or medical procedures, are available. The use of these KBs instead of general-purpose ones in such domain-specific scenarios will likely improve the accuracy and utility of the sanitized output.

Note that, in any case, the lack of a suitable knowledge base will not negatively affect the privacy protection of the output, but just the level of utility preservation, because, in the worst, case, sensitive terms will be redacted (removed) if no appropriate generalizations are available.

**A simple and scalable algorithm**

In this section we present a simple and scalable algorithm for document sanitization based on the information theoretic enforcement of our privacy model. This algorithm may serve as the basis to develop more accurate and elaborated solutions which carefully consider some of the issues discussed earlier in specific scenarios.

**Algorithm 1.**

```
Input:  D              //the input document
        KBs            //the knowledge bases
         C             //the entities to protect
         α             //the parameter of the model
Output: D'  //sanitized document

1   D'=D;
2   for each (t_i ∈ D) do
3     risky_term=false;
4     c_k=next(C);
5     while (no(risky_term) and c_k ≠ null) do
6       if (PMI(c_k;t_i)≥(IC(c_k)/α)) then //According to Definition 6
7         risky_term=true;
8       else
9         c_k=next(C);
10      end if
11    end while
12    if (risky_term) then
13      H_i=getGeneralizations(t_i, KBs); //Ordered generalizat. from KBs
14      g_t_i=next(H_i);
   //Look for a generalization that fulfills Def. 6 for all c_j in C
15      while (not (PMI(c_j;g_t_i)<(IC(c_j/α) ∀ c_j ∈ C)) do
16        g_t_i=next(H_i);
17      end while
18      sanitize(t_i, g_t_i, D'); //Replace occ. of t_i in D' by g_t_i
19    end if
20  end for
21  return D';
```

To ensure the scalability of the model implementation, the algorithm considers individual terms *t* rather than groups of terms of any cardinality. It starts by obtaining each of the different $t_i$ terms found in the input document *D* (line 2). Then, it evaluates (lines 3-11) if each $t_i$ produces a higher than desired disclosure of any entity $c_k$ in *C*, according to the instantiation of our (*α,C*)-*sanitized* model (Definition 6). If such is the case, $t_i$ is sanitized by replacing it with an appropriate generalization $g(t_i)$. To do so, an ordered list of generalizations (starting from the most specific one) is retrieved from the available knowledge bases (KBs) (line 13). Then, these generalizations are iteratively evaluated in order to discover the first one (i.e., the most specific and, thus, the most utility-preserving one) that fulfills the privacy criterion (Definition 6) for *all* the entities *C* to be protected (lines 14-17). We assume that there will always be a generalization that fulfills the privacy criterion even though it may correspond to the root concept of the KBs (e.g., *entity* in WordNet, *world* in ODP, etc.). Finally, all the occurrences of $t_i$ in the document are replaced by the suitable generalization $g(t_i)$ (line 18). When all the $t_i$ are evaluated, the resulting document *D'* will be (*α,C*)-*sanitized*.

Given that the cardinality of the entities to protect in *C* is fixed and usually small, and that the number of possible generalizations is also limited and small (e.g., a maximum of 15 for a large KB like WordNet), the algorithm solves the problem in *linear*-time with respect to the number of distinct terms $t_i$ contained the input document *D*.

## Experiments

This section presents the evaluation of a set of empirical results obtained by the implementation of the proposed model within the context of several real use cases in data privacy.

**Application examples and evaluation data**

To show the applicability of our model to the privacy requirements stated by current legislations on data privacy, we instantiated it according to the following scenarios:

- US states often dictate legislations on the privacy of medical data (Department for a Healthy New York, 2013; Health Privacy Project, 2013). These legislations mandate hospitals and healthcare organizations to redact some medical-related concepts that are considered of confidential nature before releasing patient records to, for example, insurance companies, in response to Worker's Compensation or Motor Vehicle Accident claims, or a judge, in case of malpractice litigation (Bier et al., 2009). All references to substance abuse, Sexually Transmitted Diseases (STDs) or HIV status should be usually redacted/sanitized. This redaction process, which is usually tackled manually, consumes a large amount of time and human resources, and requires of maintaining lists of terms semantically related to the confidential entities, such as medications, treatments or symptoms (Bier et al., 2009).
- US federal courts are required to redact personally identifying information about individuals from electronic or paper documents (Legal Information Institute, 2013). The same is also stated by the Health Insurance Portability and Accountability Act (HIPAA), who defines a set of elements that has to be removed from electronic healthcare records before hospitals can release them for secondary use (such as medical research). The objective is to avoid the unequivocal re-identification of identities (i.e., to preserve anonymity). Many of these elements, due to their non-semantic nature and regular structure (e.g., social security numbers, birth years, zip codes, e-mail addresses, telephone numbers etc.), can be detected by means of specifically-tailored extraction patterns and/or trained classifiers (Meystre et al., 2010), as stated above. However, there are some elements (e.g., geographical subdivisions such as counties or cities), which should be redacted from a semantic perspective, and for which the presence of semantically related terms within the document may negate its redaction.

- Data protection acts of European countries (The European Parliament and the Council of the EU, 1995) state that the information related to religion, ideology, sexual orientation or race is of sensitive nature and that companies and public organizations should protect them to avoid possible discrimination. This kind of data is specially unstructured, unbounded and inherently semantic, making it thus difficult to tackle by redaction systems based on trained classifiers.

In all these cases, the *C-sanitized* model is especially suitable, both because of the unstructured and linguistic nature of the entities to protect and also because of the need to sanitize/redact semantically-related terms.

To evaluate the implementation of our model we used, as input data, Wikipedia articles describing a set of entities considered as sensitive in the discussed use cases. As done by other related works (D. Sánchez, Batet, et al., 2013a; D. Sánchez et al., 2014; Staddon et al., 2007), Wikipedia articles were chosen due to being authoritative sources of information and also because of their high informativeness and semantically tight discourses, which configures a challenging redacting/sanitization scenario.

First, we used the Wikipedia articles corresponding to *HIV* and *STD*, which are medical conditions usually considered confidential by US states and federal laws. The instantiation of our model for each document was *HIV-sanitized*, *STD-sanitized*, respectively.

Then, we picked up geographical subdivisions corresponding to some US states with specific privacy rules on medical data (e.g., the Medical Information Reporting for California Hospitals (State of California Office of Satewide Health Planning & Development, 2013) or the New York State Confidentiality Law (Department for a Healthy New York, 2013)). Within the California state, we picked up the *Los Angeles* county Wikipedia article, in addition to the description of *New York* city. Thus, the instantiations of the proposed model were *Los Angeles-sanitized* and *New York-sanitized*, respectively.

Finally, as potentially discriminatory personal information, we used the Wikipedia articles corresponding to *Homosexuality* and *Catholicism*, for which the model instantiation were *Homosexuality-sanitized* and *Catholicism-sanitized*, respectively.

**Implementation**

The implementation of our model was based on the algorithm presented earlier. The following aspects related to the implementation and the experiments are worth noting:
- To provide a general-purpose tool, probabilities required for IC/PMI calculi were computed from the hit count provided by a Web Search Engine (Bing, in particular) when querying the evaluated terms (D. Sánchez, Batet, et al., 2013a). Thus, we are using the whole Web as an approximation of the domain knowledge *K* available to attackers.
- To mitigate the imprecisions that the Web-based probability calculus may introduce, the fact that groups of terms have not been considered in the implementation, and also to evaluate the influence of the model parameters, the (*α*,*C*)-*sanitized* model (Definition 6) was considered, with *α* values between 1.0, which corresponds to the original formulation of Definition 4, and 2.0, which results in a twice stricter sanitization, in terms of informativeness.
- SNOMED-CT, WordNet and ODP, with this priority order, were used as KBs to retrieve generalizations.
- Both redaction and sanitization were implemented in order to compare the utility gains of the latter approach when KBs are available. In the former case, lines 12-19 from Algorithm 1 were replaced by a simple removal of the terms found to be risky. In the latter case, the whole Algorithm 1 was applied.

**Evaluation criteria**

Because the main goal of our work is to be able to mimic human judgments regarding document sanitization, the evaluation has been carried out from the perspective of a human sanitizer. In this manner, we aim to evaluate whether the design of the model (unequivocal or partial disclosure of entity semantics), its information theoretic enforcement (IC as a quantification of semantics) and the specific implementation (term probabilities computed from the Web) are suitable enough to replace human sanitizers or, at least, assist them. Even though the use of a human expert focuses the evaluation on personal beliefs, at the same time, it is a realistic representation of our final goal and a good baseline, since potential attackers would hardly do better than expert sanitizers.

First, in order to evaluate the accuracy of the detection process, a human expert was requested to manually redact each of the input documents with the aim to remove terms or groups of terms that, individually or in aggregate, may help to disclose the sensitive entity *c* when publishing the documents (i.e., HIV, STD, Los Angeles, New York, Homosexuality and Catholicism). Hereinafter, we refer to the set of terms identified as sensitive by the human expert as *H*, whereas the set of terms detected by the implementation of our model is referred to as *S*. By comparing *H* and *S* sets, we can evaluate the redaction performed by our model's implementation in terms of *precision*, *recall* and *F-measure*.

*Precision* (eq. 4) measures the percentage of terms automatically detected as sensitive (*S*), which have been also identified by the human expert (*H*). The higher the precision is, the better the utility of the output because the number of non-necessary redacted terms is lower.

$$Precision = \frac{|S \cap H|}{|S|} \times 100 \qquad (4)$$

*Recall* (eq. 5) measures the percentage of correctly detected terms from the total number of terms detected by the human expert. The higher the recall is, the higher the privacy of the output will be. Recall usually plays a more important role than precision in document redaction, because a low recall may imply disclosing some data that may negate the redaction process (Anandan & Clifton, 2011).

$$Recall = \frac{|S \cap H|}{|H|} \times 100 \tag{5}$$

*F-measure* (eq. 6) measures the harmonic mean of precision and recall, thus, summarizing the accuracy of the automated redaction process.

$$F - measure = \frac{2 \times Recall \times Precision}{Recall + Precision} \tag{6}$$

To evaluate the benefits of term sanitization in front of a strict redaction, we also compared the degree of utility preservation of the output document. As stated in (Martínez et al., 2013), data semantics is the most important dimension of utility of textual data. Since we measure the semantics of terms according to their informativeness, the utility of a document *D* has been quantified as the sum of the information given by all terms appearing in *D*, as done in (D. Sánchez, Batet, et al., 2013a; D. Sánchez et al., 2014).

$$Utility(D) = \sum_{\forall t_k \in D} IC(t_k) \tag{7}$$

Then, the utility preservation of the redaction/sanitization process was measured as the ratio between the utility provided by the redacted/sanitized document *D'* with respect to the original version *D*.

$$Utility\_preservation(D') = \frac{Utility(D')}{Utility(D)} \times 100 \tag{8}$$

Notice that the sanitization performed by the human expert will tend to be the strictest possible, given her a priori knowledge on the entities to be protected and the fact that she is an expert (i.e., highly knowledgeable) in the domain. However, even though this evaluation reflects the performance of our proposal in mimicking the human sanitizer actions (which is our main goal), it does not explicitly evaluate the practical disclosure of the sanitized output (i.e., the probability that external attackers can actually discover the sensitive entities). We leave this additional evaluation as future work, which would require highly knowledgeable attackers as evaluators and a disclosure risk evaluation criteria (Martínez, Sánchez, & Valls, 2012).

**Results and discussion**

Evaluation results regarding the detection accuracy of sensitive terms for the different entities, documents and model instantiations with *α* values between 1.0 and 2.0 are summarized in Table 1.

Table 1
*Precision, recall and F-measure of the detection of sensitive terms for the different entities and model instantiations.*

| Entity/Wikipedia article | Model instantiation | Precision | Recall | F-measure |
|---|---|---|---|---|
| HIV | (1.0, HIV)-sanitized | 100% | 4% | 7,7% |
|  | (1.5, HIV)-sanitized | 96% | 88,9% | 92,3% |
|  | (2.0, HIV)-sanitized | 81,2% | 96,3% | 88,1% |
| STD | (1.0, STD)-sanitized | 100% | 8,6% | 15,8% |
|  | (1.5, STD)-sanitized | 87,5% | 60% | 71,2% |
|  | (2.0, STD)-sanitized | 87,5% | 80% | 83,6% |
| Los Angeles | (1.0, Los Angeles)-sanitized | 100% | 5,3% | 10% |
|  | (1.5, Los Angeles)-sanitized | 76,9% | 76,9% | 76,9% |
|  | (2.0, Los Angeles)-sanitized | 73,3% | 84,6% | 78,6% |
| New York | (1.0, New York)-sanitized | 100% | 8,7% | 16% |
|  | (1.5, New York)-sanitized | 71,9% | 100% | 83,6% |
|  | (2.0, New York)-sanitized | 57,1% | 100% | 72,7% |
| Homosexuality | (1.0, Homosexuality)-sanitized | 80% | 17,4% | 28,6% |
|  | (1.5, Homosexuality)-sanitized | 80,1% | 77,3% | 79,1% |
|  | (2.0, Homosexuality)-sanitized | 81,5% | 95,6% | 88% |
| Catholicism | (1.0, Catholicism)-sanitized | 100% | 8,3% | 15,4% |
|  | (1.5, Catholicism)-sanitized | 72,7% | 72,7% | 72,7% |
|  | (2.0, Catholicism)-sanitized | 69,2% | 81,8% | 75% |

Figures are quite homogenous for the different entities, although they show significant differences between each model instantiation. When $\alpha=1$, which corresponds to Definition 4, detection recall is very low in all cases (around 4-17%), whereas precision is perfect in most cases. Several causes explain this behavior. First, as discussed earlier, the web-based probability assessment may introduce imperfections in IC/PMI calculi due to language ambiguity. Concretely, the availability of different synonyms or lexicalizations of the same concept may result in an underestimation of the probability of $c$, an overestimation of its informativeness and, thus, a too loose detection process. Second, by analyzing the criterion of the human redactor, we found several cases in which non-unequivocally disclosing terms were tagged as sensitive. This shows that, in a practical scenario, disclosing a certain amount (even though not all) of the information of $c$ may be considered a privacy threat by human sanitizers. Hence, the initial model's premise could be too rigid in some practical settings. In such cases, a probable enough disclosure should be also avoided. Finally, the fact that Algorithm 1 does not consider groups of terms to ensure a good scalability results in an omission of apparently innocuous sets of terms that, when interpreted in aggregate, may cause disclosure.

To improve the detection recall, that is, the output's privacy, without compromising the implementation's efficiency, the $\alpha$ value can be increased to force a stricter assessment of the disclosure risk. Results obtained in all tests for $\alpha>1$ show a significant increase of recall (which is above 80% in all cases for $\alpha=2$) at the cost of a slightly lower precision, that is, utility. Several conclusions can be extracted from this behavior. First, the stricter criterion employed by the human expert is better approximated (in terms recall and also of global accuracy, that is, F-measure) when $α>1$. For each entity and textual resource, a particular α value would result in an optimal balance between precision and recall (i.e., highest F-measure). However, this optimal value is application-specific. Second, the lack of analysis of groups of terms can be compensated by a stricter evaluation of individual terms ($α>1$) while retaining the algorithm scalability, which solves the problem in linear time with regard to the number of terms in the document. Probability calculus inaccuracies, which may result in underestimations of disclosure risks, can also be compensated with the stricter detection criteria. Finally, according to F-measure figures, the precision decrease caused by the larger number of false positives in comparison with $α=1$, is more than compensated by the very significant increase of recall.

This shows how even a straightforward implementation of our model (Algorithm 1) can be both reasonably accurate and efficient when using the more flexible version of the proposed model (Definition 6), and how this flexibility allows implementing a stricter sanitization by raising the value of $α$ (at the cost of some precision/data utility) if achieving a high recall is critical.

Results regarding data utility preservation are shown in Table 2.

Table 2

*Data utility preservation for the different entities and model instantiations when applying term redaction (i.e., removal) or sanitization (i.e., generalization)*

| Entity/Wikipedia article | Model instantiation | Redaction | Sanitization |
|---|---|---|---|
| HIV | (1.0, HIV)-sanitized | 96,2% | 97,2% |
| | (1.5, HIV)-sanitized | 32,9% | 66,6% |
| | (2.0, HIV)-sanitized | 17,6% | 61,2% |
| STD | (1.0, STD)-sanitized | 95,3% | 97,5% |
| | (1.5, STD)-sanitized | 70,0% | 85,8% |
| | (2.0, STD)-sanitized | 65,5% | 84,5% |
| Los Angeles | (1.0, Los Angeles)-sanitized | 88,3% | 94,6% |
| | (1.5, Los Angeles)-sanitized | 54,4% | 80,1% |
| | (2.0, Los Angeles)-sanitized | 48,1% | 77,5% |
| New York | (1.0, New York)-sanitized | 97,2% | 99,2% |
| | (1.5, New York)-sanitized | 39,3% | 64,2% |
| | (2.0, New York)-sanitized | 20,1% | 58,3% |
| Homosexuality | (1.0, Homosexuality)-sanitized | 92,9% | 97,5% |
| | (1.5, Homosexuality)-sanitized | 55,3% | 81,3% |
| | (2.0, Homosexuality)-sanitized | 50,4% | 77,2% |
| Catholicism | (1.0, Catholicism)-sanitized | 96,3% | 98,1% |
| | (1.5, Catholicism)-sanitized | 41,3% | 73,4% |
| | (2.0, Catholicism)-sanitized | 30,9% | 65,8% |

First, we observe that data utility preservation is a direct function of the detection precision reported in Table 1. That is, the lower the precision is, the higher the number of false positives and, thus, the larger the amount of terms that are unnecessarily removed or generalized. Moreover, the decrease of utility preservation is sharper when redacting terms in comparison with a sanitization process. Certainly, term sanitization (i.e., generalization) enables a significant increase of data utility in all cases, which is more noticeable as the number of terms detected as sensitive increase. This shows that the exploitation of one or several knowledge bases makes the output more usable and also less obviously protected; that is, terms to be hidden are generalized instead of being blacked out, a measure that may raise awareness of their sensitivity.

In any case, the degree of preservation also depends on the generality of the concept to protect. In order to fulfill the privacy guarantee, the most general ones (e.g., New York, HIV), that is, those with a lower baseline IC, would usually require to protect a larger number of entities and to generalize them (in case of sanitization) to a higher level of abstraction than the most specific ones (e.g., Los Angeles, STD).

## Conclusions

Because of the enormous amount of unstructured sensitive documents that are daily exchanged, methods and tools aimed to automatize the burdensome sanitization/redaction process are needed. This is the purpose of the privacy model proposed in this paper, which defines the theoretical framework needed to implement practical sanitization tools that can offer clear and *a priori* privacy guarantees and an inherently semantic privacy protection that aims at mimicking the reasoning of human sanitizers. By enforcing the abstract model in terms of information theory and quantifying semantics according to the information content of terms, a practical and general purpose solution has been proposed. In comparison with other privacy models available in the literature, our proposal provides a series of benefits which include i) an intuitive and straightforward instantiation according to different privacy needs (i.e., anonymity or confidentiality) and scenarios, by using the linguistic labels usually mentioned in current privacy legislations, ii) a flexible adaptation of the degree of redaction/sanitization according to specific privacy requirements, which can be applied to individual documents, iii) a priori privacy guarantees on the kind of data protection applied to the output, iv) an automatic detection of terms that may disclose sensitive data via semantic inferences, with an inherent support of all kinds of semantic relations (taxonomic and non-taxonomic), and v) support for both redaction and sanitization (the latter requires from an appropriate KB).

As additional contributions of this work, we have i) characterized the semantics of disclosure according to the type of semantic relationship between the entities to protect and the terms appearing in the document, ii) discussed the relevant aspects that should be considered when implementing methods and tools based on our method, iii) proposed a simple and scalable implementation algorithm and iv) shown the applicability and suitability of our proposal through the evaluation, by a human expert, of the empirical results obtained in several realistic sanitization scenarios.

The proposed model and its information theoretic enforcement open the door for the development of theoretically sound redaction/sanitization methods. As future work, we plan to develop more accurate implementations of our model which would consider, for example, i) improvements on probability assessments by contextualizing web queries and, thus, minimizing semantic ambiguity inherent to raw corpora (D. Sánchez et al., 2010), ii) integration of the results provided by several web search engines in order to minimize data sparseness and compile more robust probabilities, or iii) semantic disambiguation (Roman, Hulin, Collins, & Powell, 2012) of sensitive terms in order to retrieve more appropriate generalizations. Moreover, as it has been shown through the empirical experiments, different $\alpha$ values are needed to provide an optimal balance between data privacy and utility for each specific scenario. We plan to research on ways to automate the tuning of this parameter, so that $\alpha$ can be set to an appropriate value according, for example, to the informativeness of the entity $c$ and/or of the sensitive terms found in the input document. Finally, we also plan to perform additional evaluations that consider also the perspective of potential attackers; that is, up to which point an external observer would or would not be able to infer the sensitive entities from the sanitized output.

## Disclaimer and acknowledgements

We are solely responsible for the views expressed in this paper, which do not necessarily reflect the position of UNESCO nor commit that organization. This work was partly supported by the European Commission (under projects "CLARUS" and "Inter-Trust"), by the Spanish Government (through projects ICWT TIN2012-32757 and BallotNext IPT-2012-0603-430000) and by the Government of Catalonia (under grant 2014 SGR 537). This work was also made possible through the support of a grant from Templeton World Charity Foundation. The opinions expressed in this paper are those of the authors and do not necessarily reflect the views of Templeton World Charity Foundation.